# DOES THE RR LYRAE VARIABLE DY AND SHOW THE BLAZHKO EFFECT?

*By Zachariah Cano and Robert Connon Smith*
*Astronomy Centre, University of Sussex*

Data taken on the University of Sussex 0.46-m telescope in 2006 and 2007 are combined with previously published data to obtain a better defined light curve for the RRab-type variable DY And, a slightly improved period of $0.6030897^{+0.0000006}_{-0.0000002}$ days and a new time of maximum. Evidence is presented that may indicate the Blazhko effect in this system. In addition, a new time of maximum has been obtained for VX Tri.

*Introduction*

The University of Sussex operates a 0.46-m telescope at the Isle of Thorns Observatory in the Ashdown Forest (longitude 0:01:16 E, latitude 51:03:23 N) which is used for final-year MPhys projects; further details of the telescope and CCD can be found in an earlier paper[1]. During 2006-07, a study was made of a number of RR Lyrae stars, including the RRab-type star DY Andromedae (J2000: 23 58 42.2, +41 29 19), which was followed up in the next observing season, enabling a good light curve and an improved period to be obtained.

*Observations and data reduction*

Observations of DY And were obtained on the night of 2006 November 4-5 (for an MPhys project[2]) and subsequently on six nights in 2007 September and October (Table I), comprising 1549 CCD images, each of relatively short exposure to avoid image smearing by the slight periodic error in the drive. Similar observations of VX Tri (J2000: 02 10 01, +32 24 11; this position was obtained by comparing the Aladin image with a finding chart, and is not quite the same as given by Vizier) were taken on the nights of 2006 November 6, December 9 and 10, and 2007 January 10 and 11.

Because the filter wheel was not working, all observations were taken in white light. The CCD is Peltier-cooled and was never warmer than –5 C. Times were taken from the computer clock, whose difference from UT was determined at the start of each night, and then corrected to give the heliocentric time of mid-exposure. Dark frames were subtracted automatically from each image to minimize the thermal background.

The raw images from the SBIG ST-7 camera were first converted to .fits format using a freeware program available from SBIG. Flat fields were then obtained and applied using the IRIS software[3] (version 5.55) to stack and average many images so as to blur out the star images, following the method used in the Sussex 2nd year astronomy laboratory[4]. The initial data reduction to obtain differential magnitudes was also carried out using IRIS. Subsequently, the entire post-flat-fielding reduction process was repeated for DY And using IRAF, and it was the IRAF results that were used in the period analysis reported here. Three reference stars were used for DY And; their positions are indicated in Fig. 1 and their identities and magnitudes are given in Table II. Errors in the final differential magnitudes were estimated as in the earlier paper[1]. Two nearby but unidentified reference stars were used for VX Tri; again, only differential magnitudes were obtained.

TABLE I

*Journal of observations for DY Andromedae*

| Start date | Start time (UT) | Finish time (UT) | No. of images | Exposure length (sec) |
|---|---|---|---|---|
| 2006 Nov 04 | 2246 | 0042 | 105 | 45 |
| 2006 Nov 05 | 0043 | 0345 | 158 | 30 |
| 2007 Sep 10 | 2127 | 2259 | 42 | 40 |
| 2007 Sep 11 | 2057 | 0358 | 546 | 40 |
| 2007 Sep 12 | 2043 | 2200 | 99 | 40 |
| 2007 Oct 04 | 2055 | 2201 | 89 | 40 |
| 2007 Oct 05 | 2005 | 0257 | 296 | 40 |
| 2007 Oct 14 | 2027 | 2309 | 214 | 40 |

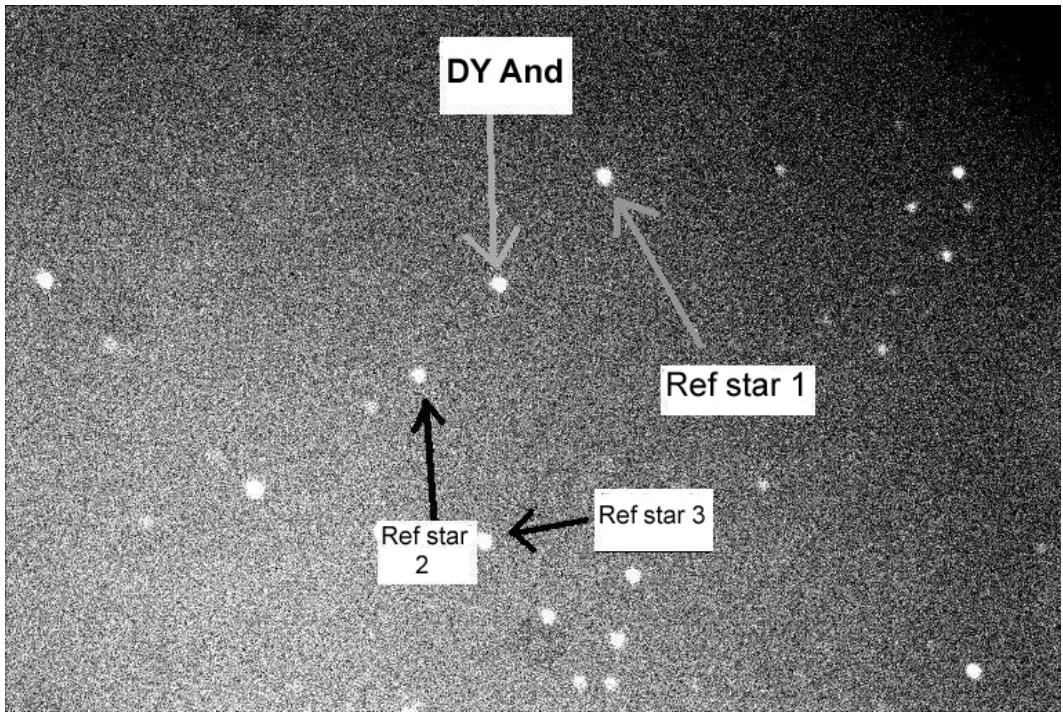

FIG. 1

A CCD image of DY And, taken on 2006 September 4, showing the reference stars used to obtain the differential magnitudes. Star 1 was the main reference star, with stars 2 and 3 primarily used as check stars (star 1 has a very close companion, whose magnitude cannot be measured separately, but both stars were included in the aperture when doing the photometry).

*Previous work on DY And*

The star was picked out as a potential target because the period given in the 1971 edition of the GCVS was not very well determined at 0.604 days (no error stated). However, later papers gave a variety of other estimates: 0.6030± 0.001 days[5], 0.603087± 0.000005 days[6] and 0.60298 days (no error quoted)[7]. The aim of the observations was either to confirm or to improve on the most precise of those periods.

TABLE II

*Data for reference stars for DY And*

|  | GSC number | Magnitude * | Comments |
|---|---|---|---|
| Ref. star 1 | 3241-0174 | 13.51 | Blended image |
| Ref. star 2 | 3241-0490 | 14.81 |  |
| Ref. star 3 | 3241-0054 | 14.69 |  |

*\* Magnitudes taken from Aladin image, on Bj system. However, only differential magnitudes were used, since our data were unfiltered.*

*Results for DY And*

The data obtained in November 2006 showed almost no variation over the five hours of observation, so it was necessary to supplement them with additional observations, obtained in September and October 2007. A period analysis was carried out using the 'string-length' option in the Starlink package PERIOD[8] and the 2007 data alone yielded a period of 0.6030 ± 0.0001 days, consistent with most previous estimates. We were also able to find a new time of maximum, at HJD 2454355.5732 ± 0.0001 days. However, a light curve for the combined 2006 and 2007 data, phased on the period of 0.6030 days, showed that the 2006 data did not fit the curve. We therefore reanalysed the combined data set and found the slightly longer period of 0.603189 days; neighbouring periods were also present, with a frequency spacing corresponding to the roughly 11-month gap between the data sets, but a light curve with this period fitted both data sets well. To resolve the difference between this period and the one published by Schmidt[6], we needed additional data.

Dr Chris Lloyd (Open University) kindly made available to us the ROTSE-I data used by Wils et al.[7], and 33 points from the Schmidt data[6] were available as an online table. Since the Sussex data were available only as differential magnitudes, we first had to add a constant to all the Sussex magnitudes to bring them onto a scale compatible with the ROTSE-I data. There was some uncertainty about this because there was a larger range in the ROTSE-I data; we chose to make the maxima have the same magnitudes. Adding the ROTSE-I data alone also gave a slightly longer period, of 0.603219 days, but in this case there was also a shorter period of nearly the same significance, namely 0.603091 days, close to Schmidt's period. The frequency difference between the two periods corresponds to a timescale of about 2800 days, which is approximately the time interval between the ROTSE-I data and the Sussex 2007 data. This suggests that the longer period is actually an alias of the true period, perhaps introduced by the small 2006 data set.

We therefore combined all three data sets, making a shift (by eye) of 0.2 in the magnitude scale to bring the unfiltered ROTSE-I and Sussex data onto a scale compatible with Schmidt's $V$ data. Analysing just the published Schmidt data[6] first, as a check, led to a period consistent with, but very slightly different from, his published value[6]; the slight difference probably arose because he also included 16 earlier data points[5], not available online. We then analysed the combined data set (Schmidt + ROTSE-I + Sussex) and found a well-defined period of 0.6030897 days; we estimate the uncertainties to be +0.0000006 and -0.0000002 days, based on the shape and width of the asymmetric minimum in the string-length/frequency plot. The nearest alternative period, considerably less significant, was again 0.603219 days, supporting the idea that this longer period is an alias.

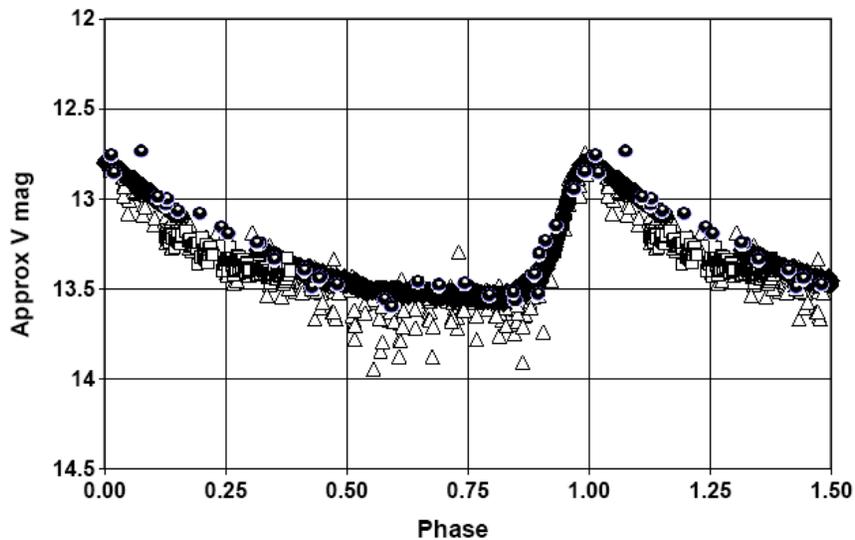

Fig. 2

The light-curve of DY And, folded on a period of 0.6030897 days. Filled circles with a white spot are data from Schmidt[6], open triangles are the ROTSE-I data, open squares are the 2006 Sussex data and filled diamonds are the 2007 Sussex data. By eye, the asymmetry in the light curve is 0.15 ± 0.05, confirming the RRab classification[5, 6].

A combined light-curve with our new period is shown in Fig. 2. The 2007 Sussex data (the filled diamonds) provide the best-defined light-curve so far published. There are three features of interest in the combined curve:

1. There is a reasonably well-defined curve for all the data, except that the 2006 Sussex data do not fit the mean curve. Since they were obtained with the same instrumental set-up as the 2007 data, and have been reduced in the same way, the mis-fit is real, suggesting a possible change in the light curve on a timescale of a year.
2. There appear to be two branches of the curve in the phase range 0.75 to 0.9; the two branches correspond to the 2007 September and 2007 October data, as can be seen in the detailed plot in Fig. 3. This may be evidence for a small change in the light curve on a timescale of 3-4 weeks.
3. There is a great deal of scatter in the ROTSE-I data, which were taken over an interval of about 8 months; this may suggest changes in the light curve on an intermediate timescale of months, although ROTSE-I data for other stars show a similar scatter.

Taken together, these three features suggest that this variable may be an unrecognised member of the class of RR Lyrae stars that show the Blazhko effect. Against this is the fact that Schmidt[6] looked for the Blazhko effect in stars in his survey, found it in some of the stars, but found no evidence of it in DY And; our suggestion is therefore rather tentative and intended mainly to flag the star as worth monitoring.

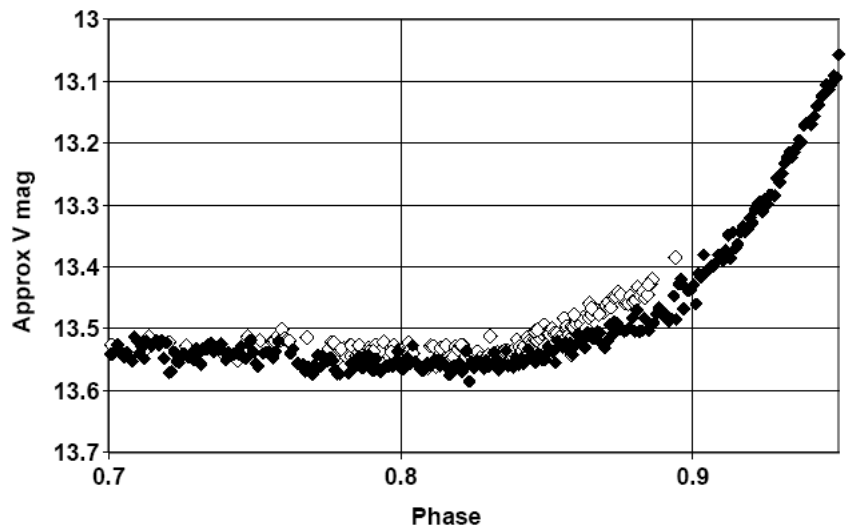

Fig. 3

Detail of Fig.2, for the Sussex data alone, to demonstrate the two branches of the light curve in the phase range 0.75 - 0.9. The solid diamonds are from 2007 September, and the open diamonds from 2007 October. Note that this feature is only detectable because of the large number of observations, closely spaced in time.

*Results for VX Tri*

The currently accepted period for VX Tri, given by Meinunger[9], is 0.633076 days. It had been hoped to combine the results of the 2006-07 project with those from a previous project in 2003-04 in order to improve the period. However, it became clear that the results from the earlier project were too noisy to be useful. Analysis of the 2006-07 data alone proved completely consistent with Meinunger's result, so the only new result reported here is a new time of maximum, at HJD 2454079.526 ± 0.005 days. The O-C diagram from the GEOS website[10] has been updated with this additional point and is shown in Fig. 4. There is no evidence of secular period change.

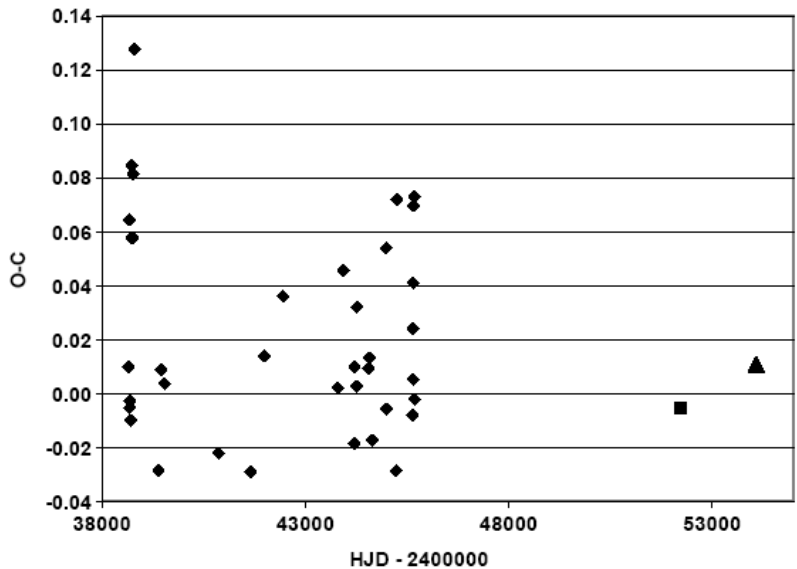

FIG. 4
The O-C diagram for VX Tri. The diamonds (from Meinunger; photographic observations) and the square (CCD observations) are taken from the GEOS database[10] and the triangle is our point.

*Conclusions*

Additional observations of DY Andromedae have produced a better light curve than previously published and, in conjunction with previous data, have enabled a slight refinement of the published period. Comparison of our data with other data available in the literature suggests that DY And may show the Blazhko effect, although there is insufficient evidence to show any periodic changes in the light curve. Clearly this is a very tentative suggestion until more data are available; we recommend further monitoring of this star.

A new time of maximum is reported for VX Tri, whose period seems to be stable.

*Acknowledgements*

The Simbad, Vizier and Aladin services of CDS (Centre de Données Astronomiques de Strasbourg) were used to obtain finding charts and other information; we also made use of the GEOS RR Lyrae Database. We thank Chris Lloyd for providing the ROTSE-I data, and a referee for helpful comments.